\documentclass[conference]{IEEEtran}
\IEEEoverridecommandlockouts
\usepackage{cite}
\usepackage{amsmath,amssymb,amsfonts}
\usepackage{algorithm}
\usepackage{algpseudocode}
\usepackage{soul}
\usepackage{tabularx}
\usepackage{multirow}
\usepackage{makecell} 
\usepackage{romannum}
\usepackage{graphicx}
\usepackage{float}
\usepackage{textcomp}
\usepackage{xcolor}
\def\BibTeX{{\rm B\kern-.05em{\sc i\kern-.025em b}\kern-.08em
    T\kern-.1667em\lower.7ex\hbox{E}\kern-.125emX}}
\begin{document}

\newcommand{\algorithmicinitialize}{\textbf{Initialize}} 
\algnewcommand{\Initialize}{\State \algorithmicinitialize\ } 
\newcommand{\algorithmicinput}{\textbf{Input:}}
\algnewcommand{\Input}{\State \algorithmicinput\ }
\newcommand{\algorithmicoutput}{\textbf{Output:}}
\algnewcommand{\Output}{\State \algorithmicoutput\ }

\title{Optimized Deployment of 
HAPS Systems for GNSS Localization Enhancement in Urban Environments\\

\thanks{This work was supported by NSERC Discovery Grant.}
}

\author{\IEEEauthorblockN{Hongzhao Zheng, Mohamed Atia, Halim Yanikomeroglu}
\IEEEauthorblockA{\textit{Department of Systems and Computer Engineering, Carleton University, Ottawa, Canada}\\
Emails: \{hongzhaozheng, halim\}@sce.carleton.ca, mohamedatia@cunet.carleton.ca}\\
}

\maketitle

\begin{abstract}
While high altitude platform stations (HAPS) have been primarily explored as network infrastructure for communication services, their advantageous characteristics also make them promising candidates for augmenting GNSS localization. This paper proposes a metaheuristic framework to jointly optimize the number and placement of HAPS for GNSS enhancement in dense urban environments, considering practical constraints such as elevation masks, altitude limits, and ray-traced visibility from 3D city models. The problem is highly nonconvex due to the discrete HAPS count and the environment-dependent 3D Cramér–Rao lower bound (CRLB). To address this, we develop a tailored version of the adaptive special-crowding distance non-dominated sorting genetic algorithm II (ASDNSGA-II). Simulations show the method successfully identifies the minimum number of HAPS needed to satisfy a CRLB threshold and selects the configuration with the lowest CRLB within that minimum, offering a cost-effective and scalable solution for future HAPS-aided positioning systems.
\end{abstract}


\section{Introduction}
Accurate outdoor localization is critical for a wide range of applications, including autonomous driving, emergency response, and augmented reality. Despite the ubiquity of global navigation satellite systems (GNSS), their accuracy is severely degraded in dense urban environments, especially in urban canyons. In these areas, tall buildings block signals, introduce multipath propagation, and create non-line-of-sight (NLOS) conditions, all of which compromise GNSS-based positioning.


To mitigate these challenges, sensor-aided localization that combines GNSS with auxiliary sensors such as light detection and ranging (LiDAR) and inertial measurement unit (IMU) have proven effective in delivering high accuracy through environmental perception and motion estimation~\cite{b1, b3}. However, these methods often rely on expensive sensors~\cite{b4}, high-quality pre-mapped environments~\cite{b5}, and are prone to drift or failure in feature-sparse areas or during long occlusions~\cite{b5}.

In contrast, while standalone GNSS or GNSS augmented with additional ranging sources typically delivers lower positioning accuracy compared to sensor-aided techniques, it offers a globally consistent and lightweight solution, making it well-suited for environments with limited infrastructure or strict cost constraints. These approaches are particularly advantageous for compact, low-power platforms such as smartphones and wearables, where GNSS capabilities are already integrated and adding extra sensors like LiDAR is often infeasible due to constraints in cost, size, energy, and processing resources. Additionally, advancements in GNSS-based localization can enhance the performance of sensor-aided systems by providing reliable priors, supporting global re-initialization, and improving resilience under degraded sensing conditions~\cite{b6,b7}.

To address these limitations while retaining the simplicity of GNSS-based approaches, high altitude platform stations (HAPS) have emerged as a promising augmentation strategy for urban localization. HAPS are envisioned as quasi-stationary aerial platforms operating in the stratosphere, typically between 18 and 22 km~\cite{b15}, offering highly favorable line-of-sight (LOS) conditions~\cite{b19,b20}. Their elevated vantage point enables extended visibility and improved spatial geometry that are difficult to achieve with terrestrial or low-altitude platforms. Prior studies have shown that integrating HAPS can significantly improve GNSS localization performance in both horizontal and vertical dimensions~\cite{b8,b9,b16}.

Nevertheless, the complex city layouts and building geometries characteristic of dense urban areas continue to pose challenges. Severe signal blockages often persist, and certain regions experience particularly poor localization performance due to enduring NLOS conditions. As a result, placement strategies based solely on geometric dilution of precision (GDOP) may overlook real-world obstructions. Although increasing the number of well-positioned ranging sources generally enhances localization accuracy, deploying an excessive number of HAPS can substantially raise operational costs and system complexity. Therefore, an effective optimization strategy that not only improves the overall localization accuracy but also minimizes the number of deployed HAPS is crucial.

Due to the discrete nature of the HAPS count and the environment-dependent objective function, jointly optimizing HAPS count and placement is inherently non-convex. Although various convexification techniques exist, they certify global optimality only when exact and may fail to capture the original problem's true feasible set. In such cases, the resulting solutions are not guaranteed to be optimal or realizable~\cite{b10}. 

As an alternative, metaheuristic methods such as genetic algorithms (GAs) are widely used to solve complex, highly non-convex problems~\cite{b11,b12,b13}. GAs are population-based search algorithms inspired by natural selection and genetics. Unlike convexification techniques that reformulate problems into convex form, GAs operate directly on the original non-convex search space without requiring convexity or differentiability. While they do not guarantee global optimality, GAs are effective in finding a high-quality near-optimal solution under restrictive assumptions.

In this paper, we propose a framework that leverages the adaptive special-crowding distance non-dominated sorting genetic algorithm (ASDNSGA-II), an enhanced version of NSGA-II known for its robustness, fast convergence, and high solution quality on benchmark multi-modal multi-objective problems (MMOPs)~\cite{b14}. The framework jointly optimizes the number and placement of HAPS to identify the minimal configuration with the least number of HAPS that satisfies a predefined 3D Cramér–Rao lower bound (CRLB) threshold, used as a quality-of-service (QoS) metric. Among all such configurations, it further selects the one achieving the lowest average CRLB over the specified regions of interest (ROI).

To account for complex urban layouts and building geometries, a 3D city model combined with ray tracing is used to evaluate the visibility between ranging sources (e.g., HAPS and satellites) and receivers. In general, the ROI consists of urban areas where localization accuracy is severely affected by obstructions such as tall buildings. Due to the high computational cost of ray tracing, especially when accounting for all possible propagation paths~\cite{b17}, evaluating the CRLB at every point is infeasible. As a scalable solution, signals are categorized as either LOS or NLOS, and the average 3D CRLB is estimated over a set of "representative locations" that capture the most challenging conditions within the ROI.

The remainder of the paper is organized as follows. Section~\Romannum{2} formulates the optimization problem. Section~\Romannum{3} presents the ASDNSGA-II algorithm and its adaptations. Section~\Romannum{4} presents the performance evaluation, including the simulation setup, parameter configurations, and simulation results. Finally, Section~\Romannum{5} concludes the paper.

\section{Problem Formulation}
\begin{figure*}[t]  
    \centering
    \includegraphics[width=0.85\textwidth, height=0.35\textheight]{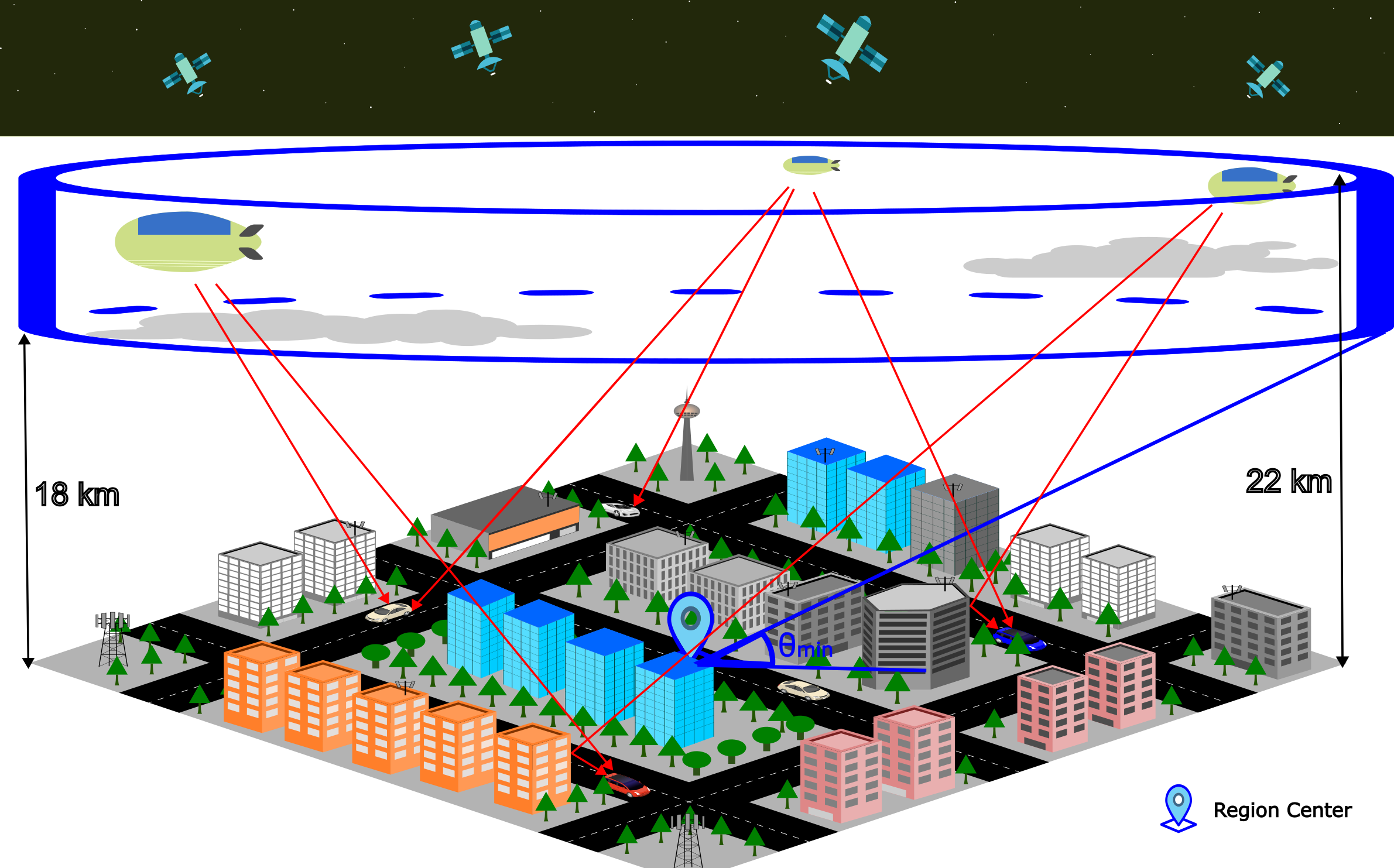}
    \caption{System model for HAPS-augmented GNSS localization in dense urban environments (HAPS are placed within a conical volume between 18km and 22km altitude while maintaining a minimum elevation angle $\theta_{\text{min}}$ relative to the region center).}
    \label{fig:system model}
\end{figure*}

The overall system model is illustrated in Fig.~\ref{fig:system model}. In this paper, we consider the Wall Street area in New York City as the dense urban region. The ROI consists of $N_{\text{r}}$ randomly selected receivers placed on the streets. Let $\mathbf{P}^{\text{r}} = [\mathbf{p}^{\text{r}}_1, \ldots, \mathbf{p}^{\text{r}}_j, \ldots, \mathbf{p}^{\text{r}}_{N_{\text{r}}}]^{\text{T}}$ denote the matrix storing all receiver positions, where $\mathbf{p}^{\text{r}}_j = [\phi^{\text{r}}_j, \lambda^{\text{r}}_j, h^{\text{r}}_j]^{\text{T}}$ represents the latitude, longitude, and altitude of receiver $j$ for all $j\in\{ 1,2,\ldots,N_{\text{r}} \}$. Similarly, let $\mathbf{P}^{\text{s}} = [\mathbf{p}^{\text{s}}_1, \mathbf{p}^{\text{s}}_2, \ldots, \mathbf{p}^{\text{s}}_{N_{\text{s},j}}]^{\text{T}}$ denote the matrix of available satellite positions for receiver $j$, where $N_{\text{s},j}$ may vary depending on the chosen elevation mask $\theta_{\text{min}}$. 

Following the general procedure of GA, each generation forms a set $\mathcal{P}_n$ consisting of $N_{\text{pop}}$ individuals. To ensure favorable geometry, the HAPS are constrained to maintain a minimum elevation angle relative to the region center $\mathbf{p}^{\text{c}}=[\phi^{\text{c}}, \lambda^{\text{c}}, h^{\text{c}}]$, and to remain within typical altitudes between 18 km and 22 km, thereby restricting their placement to a conical region in the sky. The elevation constraint is referenced to the region center, rather than individual receivers, because the ROI (Wall Street) is geographically small compared to the HAPS altitude. From such distances, the angular variation across different receivers is negligible, making the center a representative reference point. This simplification also reduces computational cost by eliminating the need for receiver-specific constraints.

The primary objective is to identify the optimal configuration of HAPS that minimizes their number while maintaining an average 3D CRLB below a predefined threshold. Furthermore, among all configurations with the same minimum number of HAPS, the one with the lowest average 3D CRLB is selected. The optimization problem can be expressed as

\begin{subequations} \label{eq:opt1}
\begin{align}
    \underset{i \in \{1,2,\ldots,N_{\text{pop}}\}}{\text{min}} & \quad N_i \label{eq:opt1_obj} \\
    \text{s.t.} \quad 
    & \frac{1}{N_\text{r}} \sum_{j=1}^{N_\text{r}} \sqrt{\operatorname{Tr}\left( \mathbf{C}^{ij}_{1:3,1:3}(\boldsymbol{\theta}^j,B) \right)} \leq \tau \label{eq:opt1_constraint}\\
    & N_i \in [N_{\text{min}}, N_{\text{max}}] \label{eq:opt1_constraint_number_of_haps}
\end{align}
\end{subequations}
\\[-4ex] 
\begin{subequations} \label{eq:opt2}
\begin{align}
    \underset{\mathbf{P}_l}{\text{min}} \quad 
    & \frac{1}{N_\text{r}} \sum_{j=1}^{N_{\text{r}}} \sqrt{\operatorname{Tr}\left( \mathbf{C}^{ij}_{1:3,1:3}(\boldsymbol{\theta}^j,B) \right)} \label{eq:opt2_obj} \\
    \text{s.t.} \quad 
    & \mathbf{P}_l \in \mathcal{P}^\star \quad \forall l \in \{1, 2, \ldots, |\mathcal{P}^\star|\}\label{eq:opt2_constraint_cardinality}\\
    \quad & \theta_{k} \geq 10^{\circ} \quad \forall k \in \{1, 2, \ldots, N_i\} \label{eq:opt2_constraint_elevation}\\
    \quad & h_k \in [h_{\text{min}},h_{\text{max}}] \label{eq:opt2_constraint_altitude}
\end{align}
\end{subequations}

\noindent where $N_i$ denotes the number of HAPS selected for the $i^{\text{th}}$ individual, and $\tau$ is the predefined CRLB threshold. $N_{\text{min}}$ and $N_{\text{max}}$ represent the predefined minimum and maximum number of HAPS, respectively. $N_{\text{pop}}$ denotes the population size of the genetic algorithm. $\mathcal{P}^\star$ denotes the set of individuals whose number of HAPS is the least among all individuals, and $|\mathcal{P}^\star|$ denotes the number of individuals in $\mathcal{P}^\star$. $\mathbf{P}_l$ denotes the $l^{\text{th}}$ individual in $\mathcal{P}^\star$, and $\theta_k$ denotes the elevation angle of the $k^{\text{th}}$ HAPS in an individual. $\mathbf{C}^{ij}_{1:3,1:3}(\boldsymbol{\theta}^j,B)$ refers to the top-left $3 \times 3$ submatrix of the CRLB matrix associated with the $i^{\text{th}}$ individual for receiver $j$, evaluated at the parameter vector $\boldsymbol{\theta}^j=[x_{\text{r}}^j,y_{\text{r}}^j,z_{\text{r}}^j,cb_{\text{r}}^j]$ and the building geometry information $B$, which is extracted from the 3D city model. $[x_{\text{r}}^j, y_{\text{r}}^j, z_{\text{r}}^j]$ denote the x-, y-, and z-coordinates of the $j^{\text{th}}$ receiver in the Earth-centered Earth-fixed (ECEF) frame, $b_{\text{r}}^j$ denotes the $j^{\text{th}}$ receiver clock bias, and $c$ is the speed of light.

The closed-form expression of the CRLB is given by
\begin{equation} 
  \mathbf{C}^{ij}(\boldsymbol{\theta},B) = \mathbf{I}^{-1}(\boldsymbol{\theta},B)
  \label{eq:crlb}
\end{equation}
\begin{equation} 
\mathbf{I}(\boldsymbol{\theta},B) = \mathbb{E} \left[ \left( \frac{\partial}{\partial \boldsymbol{\theta}} \log p(z; \boldsymbol{\theta},B) \right)
\left( \frac{\partial}{\partial \boldsymbol{\theta}} \log p(z; \boldsymbol{\theta},B) \right)^{\text{T}} \right]
\label{eq:fim_general}
\end{equation}

\noindent where $\mathbf{I}(\boldsymbol{\theta},B)$ is the Fisher information matrix (FIM) and $p(z; \boldsymbol{\theta},B)$ represents the likelihood function of observation $z$ given the parameter vector $\boldsymbol{\theta}$ and the building geometry $B$.

In dense urban environments, pseudorange residuals often exhibit complex, heavy-tailed, or multi-modal distributions due to multipath and NLOS effects, which are more accurately modeled by Gaussian mixture models (GMMs)~\cite{b18}. In this work, for LOS conditions, both satellites and HAPS are modeled using zero-mean Gaussian distributions with fixed variances, denoted as $\sigma^{\text{s}}_{\text{LOS}}$ and $\sigma^{\text{h}}_{\text{LOS}}$, respectively. For NLOS conditions, pseudorange residuals are modeled using a GMM that exclude the LOS component, with each mixture characterized by non-zero means $\mu^{\text{s}}_{\text{NLOS}}$, $\mu^{\text{h}}_{\text{NLOS}}$, variances $\sigma^{\text{s}}_{\text{NLOS}}$, $\sigma^{\text{h}}_{\text{NLOS}}$, and associated weights $w$. The parameters for satellites are chosen based on empirical residual distributions observed in the Berlin Potsdamer Platz scenario, a representative dense urban environment~\cite{b21}. 

Since HAPS pseudorange error is expected to be smaller than that for satellites~\cite{b8}, we set both means and variances for HAPS to approximately 70\% of the satellite values and use the same weights. These parameters are summarized in Table~\ref{tab:gmm_params}. Rather than drawing samples from these distributions, we compute Fisher information for each GMM component numerically using quadrature-based integration over a discretized residual domain. The resulting values are then used as inverse-variance weights in the CRLB computation.

\begin{table}[t]
\caption{Pseudorange Error Model Parameters for LOS and NLOS Conditions.}
\begin{center}
\renewcommand{\arraystretch}{1.3} 
\setlength{\tabcolsep}{4.5pt} 
\begin{tabular}{c|c|c|c|c}
\Xhline{1pt}
\textbf{Source} & \textbf{Condition} & \textbf{Mean(s) [m]} & \textbf{Std Dev(s) [m]} & \textbf{Weight(s)} \\
\Xhline{1pt}
\multirow{2}{*}{\textbf{Satellite}} 
  & LOS  & 0                      & $10$         & 1 \\
  & NLOS & $\{20, 40, 120\}$      & $\{15, 20, 50\}$                             & $\{0.5, 0.4, 0.1\}$ \\
\hline
\multirow{2}{*}{\textbf{HAPS}}     
  & LOS  & 0                      & $7$         & 1 \\
  & NLOS & $\{14, 28, 84\}$      & $\{10, 15, 35\}$                             & $\{0.5, 0.4, 0.1\}$ \\
\Xhline{1pt}
\end{tabular}
\label{tab:gmm_params}
\end{center}
\end{table}

Under the assumed measurement model, the FIM is computed by aggregating the information contribution from each individual ranging source:
\begin{equation}
\mathbf{I}(\boldsymbol{\theta},B) = \sum_{k=1}^{N} \psi_k \cdot \mathbf{h}_k^{\text{T}} \mathbf{h}_k
\label{eq:fim}
\end{equation}

\noindent where $\mathbf{h}_k$ denotes the $k$-th row of the design matrix $\mathbf{H}$ corresponding to a specific satellite or HAPS, and $\psi_k$ is the information contribution of that link. For LOS links, $\psi_k$ is set as the inverse of the assumed Gaussian variance. For NLOS links, $\psi_k$ is computed from the Fisher information of a GMM, numerically integrated over a discretized residual domain.

The joint optimization of the number and placement of HAPS remains a challenging problem due to several factors. First, the objective includes a cardinality term that penalizes the number of deployed HAPS, which is an inherently combinatorial and non-convex component. Second, the visibility conditions depend nonlinearly on HAPS placement relative to the environment. Third, the information contributions $\psi_k$ depend on the LOS/NLOS conditions, which are determined by ray-tracing through the building mesh. These factors lead to a highly non-convex, non-differentiable optimization landscape that motivates the use of evolutionary algorithms.




\section{Modified ASDNSGA-II Algorithm}
To preserve the true characteristics of the problem, we develop a modified ASDNSGA-II algorithm that directly searches for a near-optimal solution within the context of the formulated optimization problem. The core components of ASDNSGA-II include the special crowding distance (SCD) computation, an improved binary tournament selection (IBTS) strategy, and an adaptive crossover switching mechanism that switches between simulated binary crossover (SBX) and blend crossover alpha (BLX-$\alpha$).

Unlike traditional crowding distance, which measures diversity only in the objective space, SCD accounts both the decision and objective spaces to better preserve overall diversity. In IBTS, selection is primarily based on rank; when two individuals have the same rank, the one with a higher SCD score is preferred. The adaptive crossover dynamically selects between SBX and BLX-$\alpha$ by comparing the Pareto set proximity (PSP) achieved by each crossover. 

To accommodate the joint optimization of HAPS placement and count, several modifications to the original ASDNSGA-II algorithm are introduced, as detailed below.

\subsection{Modifications to Crossover and Mutation}
As the decision variables include both the number of HAPS and their spatial placement, these aspects are explicitly incorporated into the BLX-$\alpha$ and SBX crossover operations, as well as the polynomial mutation operation, to ensure the proper evolution of solutions. Additionally, rounding is applied to the resulting number of HAPS after crossover or mutation to retain its integer nature.

\subsection{Decision-Space Crowding Distance Based on ANND}
Given that individuals within the same Pareto front may have different numbers of HAPS, the standard crowding distance computation in the decision space cannot be directly applied and is therefore modified. Specifically, we adopt the aggregated nearest-neighbor distance (ANND) to fairly assess the similarity between individuals. The ANND metric calculates all pairwise distances between two individuals and aggregates the minimum distances from each point in the individual with fewer HAPS to its nearest point in the individual with an equal or greater number of HAPS.

\subsection{Heuristic Crossover Type Assignment}
Since the true Pareto set is unknown in our problem, the PSP-based crossover strategy exploited in~\cite{b14} cannot be directly applied, rendering the original ASDNSGA-II inapplicable. Therefore, we adopt a heuristic approach to select the crossover type (SBX or BLX-$\alpha$) for each individual based on its rank $r$, the crowding distances in the decision space $d_{\text{dec}}$ and objective space $d_{\text{obj}}$, as well as the predefined thresholds $d_{\text{dec}}^{\text{th}}$ and $d_{\text{obj}}^{\text{th}}$. Specifically, SBX is applied only when $r=1$, $d_{\text{dec}} > d_{\text{dec}}^{\text{th}}$, and $d_{\text{obj}} > d_{\text{obj}}^{\text{th}}$; otherwise, BLX-$\alpha$ is used.


\subsection{Modified ASDNSGA-II Procedure}
The algorithm begins by creating an initial population $\mathcal{P}_0$ consisting of $N_{\text{pop}}$ individuals, where each individual represents a candidate HAPS configuration containing $N_i$ HAPS, with $N_i \in [N_{\text{min}}, N_{\text{max}}]$. At the start of each generation, the objective matrix $\mathbf{F}_{\mathcal{P}}$, which has dimensions $N_{\text{pop}} \times 2$ and stores both the average 3D CRLB and the number of HAPS for each individual in $\mathcal{P}_{n-1}$, along with the offspring set $\mathcal{C}$ are initialized. For each individual in $\mathcal{P}_{n-1}$, the corresponding candidate solution is extracted as $\mathbf{P}_i$, and its objectives computed based on the receiver positions $\mathbf{P}^{\text{r}}$, satellite positions $\mathbf{P}^{\text{s}}$, building geometry information $B$, and noise variance $\boldsymbol{\sigma}$, where $\boldsymbol{\sigma}$ specifies the predefined pseudorange residual variances for both satellites and HAPS and for LOS and NLOS scenarios.

\begin{algorithm}[!htbp]
\caption{Modified ASDNSGA-II for HAPS Placement and Count Joint Optimization}
\label{alg:modified asdnsga ii}
\begin{algorithmic}[1]
\Input $\mathbf{P}^{\text{r}}$, $\mathbf{P}^{\text{s}}$, $\mathbf{p}^{\text{c}}$, $B$, $p_{\text{c}}$, $p_{\text{m}}$, $\boldsymbol{\sigma}$, $\eta_{\text{c}}$, $\eta_{\text{m}}$, $N_{\text{pop}}$, $N_{\text{g}}$,
\Statex \hspace{2.6em} $N_{\text{min}}$, $N_{\text{max}}$, $\theta_{\text{min}}$, $\tau$.
\Output Best HAPS configuration $\mathbf{P}^{\star}_{N_{\text{g}}}$.
\Initialize $\mathcal{P}_0$.
\For{$n = 1 : N_{\text{g}}$}
    \Initialize $\mathbf{F}_{\mathcal{P}}$, $\mathcal{C}$.
    \For{$i = 1: N_{\text{pop}}$}
        \State $\mathbf{P}_i \gets \mathcal{P}_{n-1}$. 
        \State Compute $\mathbf{F}_{\mathcal{P}}$ based on $\mathbf{P}^{\text{r}}$, $\mathbf{P}^{\text{s}}$, $B$, and $\boldsymbol{\sigma}$.
    \EndFor
    \State Compute $\mathcal{F}$ and $\mathbf{r}$ via FNS on $\mathbf{F}_{\mathcal{P}}$.
    \State Find the best solution $\mathbf{P}_n^\star$ via elite selection.
    \State Compute $\mathbf{d}_{\text{dec}}$, $\mathbf{d}_{\text{obj}}$, and $\mathbf{d}_{\text{SCD}}$ via SCD for all 
    \Statex \hspace{1.2em} individuals in $\mathcal{P}_{n-1}$ based on $\mathbf{F}_{\mathcal{P}}$ and $\mathcal{F}$.
    \State Select parents and store them in $\mathcal{P}_n$ via IBTS.
    \For{$i = 1 : 2 : N_{\text{pop}}$}
        \State $\mathbf{P}_i,\mathbf{P}_{i+1} \gets \mathcal{P}_n$.
        \State $r_i,r_{i+1} \gets \mathbf{r}$.
        \State $d_{\text{dec},i}, d_{\text{dec},{i+1}} \gets \mathbf{d}_{\text{dec}}$.
        \State $d_{\text{obj},i}, d_{\text{obj},{i+1}} \gets \mathbf{d}_{\text{obj}}$.
        \If{$rand \leq p_{\text{c}}$}
            \State {Apply the adaptive crossover with the heuristic 
            \Statex \hspace{4.2em} crossover type assignment on $\mathbf{P}_i$ and $\mathbf{P}_{i+1}$.}
        \EndIf
        \If{$rand \leq p_\text{m}$}
            \State Apply polynomial mutation on $\mathbf{P}_i$ and $\mathbf{P}_{i+1}$.
        \EndIf
        \State Add children $\mathbf{P}_i$ and $\mathbf{P}_{i+1}$ to offspring set $\mathcal{C}_{n-1}$.
    \EndFor
    \State Evaluate offspring objectives $\mathbf{F}_{\mathcal{C}}$.
    \State Combine population $\mathcal{P}_{n-1} \cup \mathcal{C}_{n-1}$.
    \State Combine objectives $\mathbf{F}_{\mathcal{P}} \cup \mathbf{F}_{\mathcal{C}}$.
    \State Get $\mathcal{C}_n$ via environmental selection on $\mathcal{P}_{n-1} \cup \mathcal{C}_{n-1}$.
    \State $\mathcal{P}_n = \mathcal{C}_n$.
\EndFor
\end{algorithmic}
\end{algorithm}

Subsequently, the fast non-dominated sorting (FNS) is performed on the objectives $\mathbf{F}_{\mathcal{P}}$ to obtain the Pareto front set $\mathcal{F}$ and the ranks $\mathbf{r}$. The elite solution $\mathbf{P}_n^\star$ is then identified through elite selection. Following this, the crowding distances in decision space $\mathbf{d}_{\text{dec}}$, objective space $\mathbf{d}_{\text{obj}}$, and the special crowding distances $\mathbf{d}_{\text{SCD}}$ are computed for all individuals, where both $\mathbf{d}_{\text{obj}}$ and $\mathbf{d}_{\text{SCD}}$ are calculated based on~\cite{b14}.

Parent selection is then performed using the IBTS from~\cite{b14}, resulting in $N_{\text{pop}}$ parents stored in $\mathcal{P}_n$. For each pair of parents, their ranks ($r_i$, $r_{i+1}$), decision-space crowding distances ($d_{\text{dec},i}$, $d_{\text{dec},{i+1}}$), and objective-space crowding distances ($d_{\text{obj},i}$, $d_{\text{obj},{i+1}}$) are retrieved. Adaptive crossover with the heuristic crossover type assignment is then applied to $\mathbf{P}_i$ and $\mathbf{P}_{i+1}$. Subsequently, polynomial mutation is applied to introduce variations. The resulting children are added to the offspring set $\mathcal{C}_{n-1}$.

After generating the offspring set $\mathcal{C}_{n-1}$, the objective values of all child individuals are evaluated in the same manner as the parent population. Environmental selection is then applied to the combined pool $\mathcal{P}_{n-1} \cup \mathcal{C}_{n-1}$, where FNS is first used to identify Pareto fronts. Complete fronts from the first one are sequentially added to the next generation, while individuals from the last partial front are selected based on their SCD, prioritizing solutions that contribute most to diversity. This process yields a new population $\mathcal{C}_n$ of size $N_{\text{pop}}$. The population is subsequently updated as $\mathcal{P}_n = \mathcal{C}_n$, marking the completion of a generation. The complete procedure of the modified ASDNSGA-II algorithm for HAPS placement and count joint optimization is shown in Alg.~\ref{alg:modified asdnsga ii}. 

Due to the characteristics of SBX, BLX-$\alpha$, and polynomial mutation, the updated HAPS locations may occasionally fall outside the confined conical region. When this occurs, the locations are projected back to the nearest valid point within the allowable region.

\section{Performance Evaluation}

\begin{figure}[t]
\centering
\includegraphics[width=0.85\columnwidth]{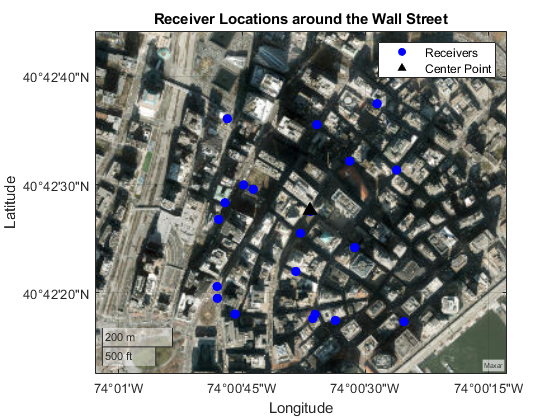}
\caption{Randomly placed receiver locations around the Wall Street.}
\label{fig:receiver locations}
\end{figure}

We utilize MATLAB’s Satellite Communications Toolbox to simulate the satellite scenario and determine available satellites for each receiver based on a given timestamp and elevation mask. To mitigate edge effects caused by incomplete building data near the boundaries of the 3D city model, which could artificially improve LOS conditions to the HAPS, receiver locations are carefully selected to be well within the interior of the model. The distribution of these receivers is illustrated in Fig.~\ref{fig:receiver locations}. With the inclusion of additional ranging sources, the elevation mask is set to 10$^\circ$ for both satellites and HAPS. The crossover probability $p_{\text{c}}$, mutation probability $p_{\text{m}}$, and the auxiliary parameters for SBX and polynomial mutation, $\eta_{\text{c}}$ and $\eta_{\text{m}}$, respectively, are configured following the settings reported in~\cite{b14}. Considering the dense urban environment analyzed in this study, the CRLB threshold $\tau$ is set to 20 m. A summary of all simulation parameters is provided in Table~\ref{tab:simulation parameters}.

\begin{table}[t]
\caption{Simulation parameters.}
\begin{center}
\renewcommand{\arraystretch}{1.3} 
\setlength{\tabcolsep}{6pt} 
\begin{tabular}{>{\centering\arraybackslash}m{0.15\linewidth}|>{\centering\arraybackslash}m{0.15\linewidth}||>{\centering\arraybackslash}m{0.15\linewidth}|>{\centering\arraybackslash}m{0.15\linewidth}}
\Xhline{1pt}
\textbf{Parameter} & \textbf{Value} & \textbf{Parameter} & \textbf{Value} \\
\Xhline{1pt}
$p_{\text{c}}$ & 0.9 & $p_{\text{m}}$ & 0.01 \\
\hline
$\eta_{\text{c}}$ & 20 & $\eta_{\text{m}}$ & 20 \\
\hline
$N_{\text{pop}}$ & 50 & $N_{\text{g}}$ & 100 \\
\hline
$N_{\text{min}}$ & 1 & $N_{\text{max}}$ & 8 \\
\hline
$\theta_{\text{min}}$ & 10$^\circ$ & $\tau$ & 20 m \\
\hline
$d_{\text{dec}}^{\text{th}}$ & 0.5 & $d_{\text{obj}}^{\text{th}}$ & 0.5 \\
\Xhline{1pt}
\end{tabular}
\label{tab:simulation parameters}
\end{center}
\end{table}

\begin{figure}[!t]
\centering
\includegraphics[width=0.85\columnwidth]{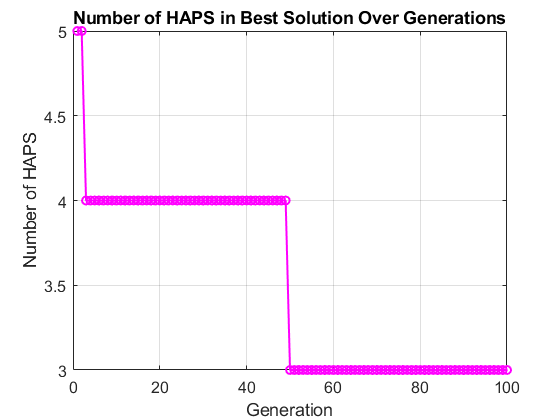}
\caption{Number of HAPS in best solution over generations.}
\label{fig:number of haps in best solution over generations}
\end{figure}

Based on the specified simulation settings, we evaluate the performance of the modified ASDNSGA-II algorithm in identifying near-optimal solutions for both HAPS placement and count. Fig.~\ref{fig:number of haps in best solution over generations} illustrates the number of HAPS in the best solution found across generations. As expected, this number is monotonically non-increasing, indicating that the algorithm progressively identifies more efficient configurations.

\begin{figure}[!t]
\centering
\includegraphics[width=0.85\columnwidth]{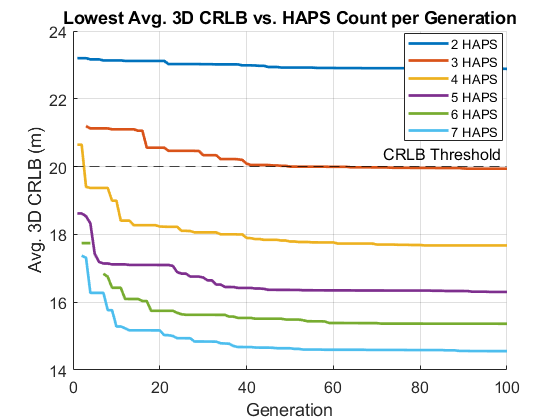}
\caption{Lowest average 3D CRLB vs HAPS count per generation.}
\label{fig:best avg 3d crlb vs haps count per generation}
\end{figure}

Fig.~\ref{fig:best avg 3d crlb vs haps count per generation} shows the lowest average 3D CRLB for each HAPS count across generations. The values steadily decrease, demonstrating the algorithm's effectiveness in improving localization accuracy over time. Higher HAPS counts (e.g., 5–7) consistently yield lower CRLBs, while the 2-HAPS case remains above the threshold, indicating insufficient accuracy. Notably, the 3-HAPS configuration eventually falls below the threshold, revealing cost-effective solutions that balance deployment cost and performance. These results confirm the algorithm’s ability to converge toward optimal or near-optimal configurations while satisfying both accuracy and resource constraints.

\begin{figure}[!t]
\centering
\includegraphics[width=0.85\columnwidth]{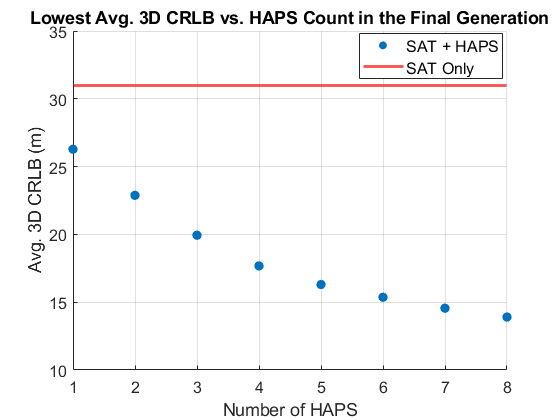}
\caption{Lowest average 3D CRLB vs HAPS count in the final generation (blue dots: lowest average 3D CRLB; red line: average 3D CRLB).}
\label{fig:best avg 3d crlb vs haps count_final}
\end{figure}

Lastly, Fig.~\ref{fig:best avg 3d crlb vs haps count_final} presents the lowest average 3D CRLB achieved for each distinct number of HAPS in the final generation, alongside the average 3D CRLB for the satellite-only case as a baseline. It is observed that introducing the first HAPS reduces the CRLB from approximately 31 m to 26.3 m, and a second HAPS further lowers it to about 22.9 m, demonstrating substantial initial gains. However, the improvement per additional HAPS diminishes rapidly; for instance, increasing from four to five HAPS only reduces the CRLB from around 17.7 m to 16.3 m. Beyond five or six HAPS, the gains become marginal, indicating a saturation effect. These results suggest that deploying four HAPS provides an effective trade-off between localization accuracy and deployment cost.

\section{Conclusion}
The simulation results demonstrate that the modified ASDNSGA-II algorithm effectively addresses the joint optimization of HAPS number and placement for GNSS enhancement in dense urban environments. The algorithm consistently identifies the minimum number of HAPS required to satisfy stringent CRLB thresholds and, within this minimum, selects the configuration that achieves the lowest average 3D CRLB. By incorporating constraints such as elevation masks, altitude ranges, and LOS/NLOS conditions derived from 3D city-model ray tracing, the framework ensures realistic applicability. These findings confirm that the proposed approach enhances localization accuracy while minimizing infrastructure deployment, offering a cost-effective and scalable solution for high-precision GNSS augmentation with HAPS.




\begin{thebibliography}{00}
\bibitem{b1} L. Li, M. Yang, H. Li, C. Wang, and B. Wang, ``Robust localization for intelligent vehicles based on compressed road scene map in urban environments,'' \textit{IEEE Trans. Intell. Veh.}, vol. 8, no. 1, pp. 250-262, Jan. 2023.



\bibitem{b3} J. Liu, Y. Liang, D. Xu, X. Gong, and J. Hyyppä, ``A ubiquitous positioning solution of integrating GNSS with LiDAR odometry and 3D map for autonomous driving in urban environments,'' \textit{J. Geod.}, vol. 97, no. 39, Apr. 2023.

\bibitem{b4} Y. Cai, Z. Lu, H. Wang, L. Chen, and Y. Li, ``A lightweight feature map creation method for intelligent vehicle localization in urban road environments,'' \textit{IEEE Trans. Instrum. Meas.}, vol. 71, pp. 1-15, Jun. 2022.

\bibitem{b5} T. Tuna, J. Nubert, Y. Nava, S. Khattak, and M. Hutter, ``X-ICP: Localizability-aware LiDAR registration for robust localization in extreme environments,'' \textit{IEEE Trans. Robot.}, vol. 40, pp. 452-471, Nov. 2024.

\bibitem{b6} B. Han, Z. Xiao, S. Huang, and T. Zhang, ``Multi-layer VI-GNSS global positioning framework with numerical solution aided MAP initialization,'' in \textit{Proc. IEEE/RSJ Int. Conf. Intell. Robots Syst. (IROS)}, Prague, Czech Republic, Sep. 2021, pp. 5448-5455.

\bibitem{b7} Y. Luo, L. Hsu, Y. Jiang, B. Liu, Z. Zhang, Y. Xiang, and N. El-Sheimy, ``High-accuracy absolute-position-aided code phase tracking based on RTK/INS deep integration in challenging static scenarios,'' \textit{Remote Sens.}, vol. 15, no. 4, pp. 1114, Feb. 2023.

\bibitem{b15} S. C. Arum, D. Grace, and P. D. Mitchell, ``A review of wireless communication using high-altitude platforms for extended coverage and capacity,'' \textit{Comput. Commun.}, vol. 157, pp. 232-256, May 2022.

\bibitem{b19} G. Karabulut Kurt et al., ``A vision and framework for the high altitude platform station (HAPS) networks of the future,'' \textit{IEEE Communications Surveys \& Tutorials}, vol. 23, no. 2, pp. 729-779, Secondquarter 2021.

\bibitem{b20} S. Alfattani, W. Jaafar, Y. Hmamouche, H. Yanikomeroglu, and A. Yongaçoglu, ``Link budget analysis for reconfigurable smart surfaces in aerial platforms,'' \textit{IEEE Open J. Commun. Soc.}, vol. 2, pp. 1980-1995, 2021.

\bibitem{b8} H. Zheng, M. Atia, and H. Yanikomeroglu, ``Analysis of a HAPS-aided GNSS in urban areas using a RAIM algorithm," \textit{IEEE Open J. Commun. Soc.}, vol. 4, pp. 226-238, 2023.

\bibitem{b9} T. Tsujii, M. Harigae, and K. Okano, ``A new positioning/navigation system based on pseudolites installed on high altitude platforms systems (HAPS),'' in \textit{Proc. 24th Int. Congr. Aeronaut. Sci. (ICAS)}, Yokohama, Japan, Sep. 2004, pp. 32.


\bibitem{b16} H. Zheng, M. Atia, and H. Yanikomeroglu, ``A positioning system in an urban vertical heterogeneous network (VHetNet),'' \textit{IEEE J. Radio Freq. Identif.}, vol. 7, no. 390-401, Apr. 2023.

\bibitem{b10} S. H. Low, ``Convex relaxation of optimal power flow—part I: Formulations and equivalence," \textit{IEEE Trans. Control Netw. Syst.}, vol. 1, no. 1, pp. 15-27, Mar. 2014.

\bibitem{b11} M. Sohrabi, A. M. Fathollahi-Fard, and V. A. Gromov, ``Genetic engineering algorithm (GEA): An efficient metaheuristic algorithm for solving combinatorial optimization problems,'' \textit{Autom. Remote Control}, vol. 85, no. 3, pp. 252-262, Aug. 2024.

\bibitem{b12} N. Demo, M. Tezzele, and G. Rozza, ``A supervised learning approach involving active subspaces for an efficient genetic algorithm in high-dimensional optimization problems,'' \textit{SIAM J. Sci. Comput.}, vol. 43, no. 3, pp. B831-B853, 2021.

\bibitem{b13} B. Tomoiagă, M. Chindriş, A. Sumper, A. Sudria-Andreu, and R. Villafafila-Robles, ``Pareto optimal reconfiguration of power distribution systems using a genetic algorithm based on NSGA-II,'' \textit{Energies}, vol. 6, no. 3, pp. 1439-1455, Mar. 2013.

\bibitem{b14} W. Deng, X. Zhang, Y. Zhou, Y. Liu, X. Zhou, H. Chen, and H. Zhao, ``An enhanced fast non-dominated solution sorting genetic algorithm for multi-objective problems,'' \textit{Inf. Sci.}, vol. 585, pp. 441-453, Mar. 2022.


\bibitem{b17} H. Zheng, M. Atia, and H. Yanikomeroglu, ``Realistic channel and delay coefficient generation for dual mobile space-ground links: A tutorial,'' \textit{IEEE Open J. Veh. Technol.}, vol. 5, pp. 762-777, 2024.

\bibitem{b18} N. Viandier, D. F. Nahimana, J. Marais, and E. Duflos, ``GNSS performance enhancement in urban environment based on pseudo-range error model,'' in \textit{Proc. IEEE/ION Position, Location and Navigation Symposium (PLANS)}, Monterey, USA, May 2008, pp. 377-382.

\bibitem{b21} T. Pfeifer and P. Protzel, ``Expectation-maximization for adaptive mixture models in graph optimization,'' in \textit{Proc. IEEE Int. Conf. Robot. Autom. (ICRA)}, Montreal, Canada, May 2019, pp. 3151-3157.










\end{thebibliography}
\end{document}